\documentstyle[12pt]{article}
\begin{document}
\title{The Ngai model of relaxation: An alternative for the study
of an interacting system of small particles}
\author{
Running Title: Relaxation in an interacting fine\\
 particle assembly\\
Subject Classification: 75.60\\
J. Castro\\
Departamento de Fisica Aplicada\\
Universidad de Santiago de Compostela\\
E-15706, Spain}
\maketitle
\begin{abstract}
It is shown that the Ngai model of relaxation
 accounts for the main experimental results
of the magnetic relaxation in an assembly of interacting fine particles.

\end{abstract}

\newpage
\section{Introduction}
The Ngai model for relaxation in complex systems was first introduced in
1979 \cite{uno,dos}. Since then,
it has been found to offer an accurate description of a wide variety
of systems~\cite{tres}. However, its use in magnetism is more recent.
See reference~\cite{cuatro} and other references therein

The fundamental difference between the Ngai model and
previous models of relaxation is that these  earlier models treat the
effect of complexity on the relaxation as being essentially static, in the
sense that the effect of complexity is to change the relaxation
time or to produce the presence of a distribution of relaxation
times. In contrast to this, in the Ngai model, the relaxation in complex
systems, is dynamical in nature.

From this view-point, according to Ngai~\cite{cinco}, a relaxing complex
system consists of three parts:

i) Individual primary species (PS) of which are of interest for the relaxation.

ii) A heat-bath whose interaction with the PS provides a primary
mechanism of relaxation.

iii) Other relaxing species, X, whose interaction with PS, the PS-X
coupling, slows down the relaxation process. This is the main
manifestation of complexity.

In this paper we propose another application of the Ngai model to
a magnetic system: namely that of an interacting fine-particle 
assembly. In this system, the PS are the individual particles 
whose coupling with the other particles in the assembly slows 
down the relaxation process. In preceding attempts to tackle 
this problem \cite{seis,siete},
the only effect of the presence of interparticle interactions is 
a change in the relaxation times. A discussion of these results 
in the light of the present model, together an evaluation of experimental
results using this model, are also presented.

In this paper we show that many of the experimental results for an
interacting system of small particles come out naturally within
the Ngai model. In particular the model can explain,
in a natural way, not only the abserved differences for a system
of interacting and non-interacting particles via the variation of
the inverse of the blocking temperature with the logarithm of the
measuring time, but also the 
values encountered for the relaxation times (table 1).  The
organization of the paper is as follows: in section 2 we review the
Ngai model for the relaxation in a complex system. In section 3 we
apply this model to a system of interacting particles and show that
the behaviour  discussed previously can be accounted for by this
model. Section 4 is  devoted to the calculus of the magnetic
viscosity in two different extentions of the model, the case of a
distribution of particle-sizes and also the case of a distribution
of switching fields. In the former, we encounter the ubiquitous
logarithmic relaxation law, whereas in the latter our results
agree  with  those of Charap\cite{charap} and therefore the model
is compatible with the experimental results of Oseroff et
al\cite{oseroff}.

\section{The Ngai model}
After the justification of the necesity of the study of 
non simple-exponential relaxations we sumarize briefly the more
important results of the Ngai model . For more
information about the theoretical derivations and the
experimental succeses of the model we refer to the
bibliography~\cite{cinco}.

It was shown by Ngai et al~\cite{nrrt} that all the permissible
relaxation functions have to verify the Paley-Wiener theorem of the
Fourier transform theory. It is easy to show that the simple
exponential relaxation function violates this theorem. Sice a
single, linear exponential form is unphysical, as demonstrated,
the idea of a superposition of exponentially decaying functions
may be also ruled out as a description of relaxation phenomena.
There is therefore the neccessity to move on non
simple-exponential relaxations.

The Ngai model~\cite{cinco} proposes the existence of a temperature-insensitive
crossover time, $t_{c}$, separating two regimes in which the dynamics
of the relaxation are different. The relaxation function is a
linear exponential $exp(-\frac{t}{\tau_{0}})$ for $t<t_{c}$ and
cross-over at $t\approx t_{c}$ to a stretched exponential form
$exp(-(\frac{t}{\tau_{\star}})^{(1-n)})$ for $t>t_{c}$. The coupling
parameter $n$ lies in the range $0<n<1$. Notice the appearance of
two different characteristic time-scales, $\tau_{0}$ and
$\tau_{\star}$, for the two regimes. These two time-scales are not
independent, but continuity of the relaxation function at $t=t_{c}$
implies that:
\begin{equation}
\tau_{\star}=(t_{c}^{(-n)}\tau_{0})^{\frac{1}{1-n}}
\label{a}
\end{equation}
Notice an important consequence of equation~\ref{a} which will be
important in the following, if $\tau_{0}$ describes the physics of
some motion over an energy barrier with activation
energy $E_{A}$:
\begin{equation}
\tau_{0}=\tau_{\infty}exp(\frac{E_{A}}{kT})
\end{equation}

Then for equation~\ref{a}, $\tau_{\star}$ describes a macroscopic
relaxation time given by:
\begin{equation}
\tau_{\star}=\tau_{\infty}^{\star}exp(\frac{E_{A}^{\star}}{kT})
\end{equation}
where:
\begin{equation}
E_{A}^{\star}=\frac{E_{A}}{1-n}
\end{equation}
and:
\begin{equation}
\tau_{\infty}^{\star}=\tau_{\infty}^{\frac{1}{1-n}}t_{c}^{\frac{n}{n-1}}
\end{equation}
Where $E_{A}^{\star}>E_{A}$.
\section{Application of the Ngai model to the relaxation of a system of
 small interacting
particles}
There are several excellents treatment of the magnetic properties of
an ensemble of particles. We shall, therefore, not enter into
detail here
\cite{kneller,chantrell}.

We consider, firstly, an ensemble of particles of equal volume $V$.
For simplicity we assume that the particle anisotropy is uniaxial and that
the particle's easy axis of magnetization is parallel to the field direction.
Then the energy of one of the particles in a field $H$ will be \cite{ocho}:
\begin{equation}
E=KVsin^{2}\theta-M_{s}VHcos\theta
\end{equation}
where $K$ is the anisotropy constant, $M_{s}$ is the bulk 
saturation magnetization
and $\theta$ measures the orientation of the magnetic moment of the
particle with respect to the field. This
energy gives rise to  two energy minima separated by an energy barrier:
\begin{equation}
E_{A}=\frac{V}{2}M_{s}H_{K}(1-\frac{H}{H_{K}})^{2}=e_{A}V
\end{equation}
At a finite temperature, we have a finite probability of a thermally
activated motion over the energy barrier. The characteristic 
time-scale of the motion will be given by the expression:
\begin{equation}
\tau_{0}=\tau_{\infty}exp(E_{A}/kT)
\end{equation}
where $\tau_{\infty}\approx10^{-9}-10^{-12}$s \cite{nueve}.
For such a system, the Ngai model predicts the existence of a cross-over time,
$t_{c}$, independent of  temperature,  to be determined from the experiment,
separating two different regimes of the dynamics of the relaxation. 
The relaxation
function is a linear exponential $exp(-t/\tau_{0})$ for $t<t_{c}$ and
has a stretched exponential form $exp(-(t/\tau_{\star})^{1-n})$ for $t>t_{c}$,
where $0<n<1$. The relation between $\tau_{\star}$ and $\tau_{0}$ is given
by equation 1.

An important consequence of the model can be now pointed out if,
as a definition of the blocking temperature, we use the usual
criterion that the characteristic time-scales of the relaxation
will be of the order of the measuring time. We then expect that the
logarithm of the measuring time plotted as a function of
$1/T_{B}$ will be a \lq\lq piece-wise" straight line with two different
slopes, $\frac{E_{A}}{k_{B}}$ for $t<t_{c}$ and $\frac{E_{A}}{k_{B}(1-n)}$ for
$t>t_{c}$. The slope corresponding to the small measuring times will be 
smaller than that corresponds to the longer measuring times.
This fact was encountered experimentally in a detailed analysis of the
dynamics of a system of small iron particles, magnetically interacting and
dispersed in an amorphous alumina matrix by 
Dormann  et al \cite{siete}.
We reproduce here data from a figure from that paper, figure 2, which shows the predicted
behaviour (figure 1). From this figure, we estimate that $t_{c}$ is in the range
$10^{-10}-10^{-5}s$. However these results are not immediately comparable with
the above developed theory because they correspond to samples with different
sizes and interactions.

A better comparison with the theory can be made with the data of 
J. L. Dormann et al\cite{mio} in a system consisting of $\gamma-Fe_{2}O_{3}$
particles dispersed in polyvinilic alcohol (figure 2)
\footnote{I am indebted to Dr. D. Fiorani for pointing this
possibility  out to me
and for making available his experimental data}. In this case we have 
a series of samples
with the same size and different interactions including the case of zero
interaction (sample IF). The interaction increases from right 
to left. From the
sample with zero interaction we can estimate $\tau_{\infty}\approx10^{-10}s$.
By comparing the slopes of sample IF with the slopes of the other samples,
we can calculate the value of $n$. Using these values of $n$, 
the afore-mentioned
value of $\tau_{\infty}$ and choosing $t_{c}\approx10^{-9}$, we can
calculate theoretically the value of $\tau_{\infty}^{\star}$. These values are
compared with the experimentally obtained values in table 1. We observe 
excellent agreement between theory and experiment.

\begin{figure}
\vspace{6in}
\caption{Variation of the blocking temperature, $T_{B}$, with the logarithm
of the measuring time for different samples of iron particles,
 according to 
[7]. The average particle size and the
interparticle interaction strength 
increases on going from  right  to left.}
\end{figure}

\begin{figure}
\vspace{6in}
\caption{Variation of the blocking temperature, $T_{B}$, with the logarithm
of the measuring time for different samples of
$\gamma-Fe_{2}O_{3}$ particles, according to [14]. The interparticle
interaction strength increases from right to left.  The average
particle size is held constant.} \end{figure}

\begin{table}
\begin{center}
\begin{tabular}{|c|c|c|c|}
\hline
Sample&n&$\tau_{\infty}^{\star}(th)(s)$&$\tau_{\infty}^{\star}(exp)(s)$\\
\hline
CH&0.4&$4\times 10^{-12}$&$ 10^{-12}$\\
\hline
IN&0.73&$3.87\times 10^{-16}$&$ 10^{-16}$\\
\hline
FLOC&0.76&$3\times 10^{-17}$&$ 10^{-17}$\\
\hline
\end{tabular}
\end{center}
\caption{Comparison between the theoretical and experimentally obtained values
for $\tau_{\infty}^{\star}$. See text.}
\end{table}

Here we wish to stress the importance of this fact. The existence of the
cross-over time, which separates two different time dependences with 
times-scales related by equation 5, is a stringent condition 
for the verification of the Ngai model. This is the first time that 
this cross-over has been encountered in a magnetic system.

A further important consequence is that the Ngai theory can explain the
discrepancy between the size of the particles measured using two different
time scales. This will be investigated in a further publication\cite{castro}.

After showing that the Ngai can describe
 the relaxation of a system of small interacting particles we dedicate
the remainder of this paper to the extention of the model to a 
calculation of the magnetic viscosity in the case of a distribution 
of particle sizes and a distribution of switching fields.
\section{Calculation of the magnetic viscosity}
A typical viscosity measurement in a small particle assembly runs as follows:
After saturation in a strong field, a small measuring field   is applied in the
reverse direction. The magnetization is then measured as a function of
 time. Typical times between application of the saturation and 
 the first measurement are of the order of seconds and the 
 time-scale of the measurement is of
the order of minutes. In the previous section, we estimate that
$t_{c}\approx10^{-10}-10^{-5}s$ (figure 1) and 
thus we expect that the relaxation
function for the present experiment will be of the stretched exponential type.
Using the boundary conditions, $M(t=0)=-M_{0}$ and $M(t=\infty)=M_{0}$
the time dependence of the magnetization can be written in the form:
\begin{equation}
M(t)=M_{0}-2M_{0}exp\{-(\frac{t}{\tau_{\star}})^{1-n}\}
\end{equation}
We now extend the model to two cases: distribution of particle sizes
and distribution of switching fields.
\subsection{Distribution in the particle size}
We now consider the fact that we have a distribution in the particle size,
$f(V)$ such that:
\begin{equation}
\int_{0}^{\infty}f(V)dV=1
\end{equation}
In this caes, our relaxation law transforms to:
\begin{equation}
\frac{M(t)}{M_{0}}=\int_{0}^{\infty}(1-2exp\{-(\frac{t}{\tau_{\star}})^{1-n}\})f(V)dV
\end{equation}
As a  simple example we consider the distribution \cite{diez}:
\begin{eqnarray}
f(V)=\left\{
\begin{array}{ll}
\frac{1}{V_{2}-V_{1}} & \mbox{$V_{1}<V<V_{2}$} \\
0 & \mbox{otherwise}
\end{array}
\right.
\end{eqnarray}

For this distribution function, the relaxation function can be evaluated
in terms of the exponential integral:
\begin{equation}
E_{i}(x)=\int_{x}^{\infty}\frac{e^{-y}}{y}dy
\end{equation}
A simple calculation gives:
\begin{equation}
M(t)=M_{0}-2M_{0}\frac{kT}{e_{A}(V_{2}-V_{1})}(E_{i}(z_{2})-E_{i}(z_{1}))
\end{equation}
where:
\begin{equation}
z_{i}=(\frac{t}{\tau_{\infty}^{\star}exp(\frac{e_{A}V_{i}}{kT})})^{1-n}\hspace{0.2in}i=1,2
\end{equation}
We stress here that this opens up a new possibility for the analysis of the
 relaxation data.

For $V_{1}$, $V_{2}$, $n$ and $t$ such that $z_{1}>>1$ and $z_{2}<<1$ we have:
\begin{equation}
M(t)=M_{0}-\frac{2M_{0}k_{B}T}{e_{A}(V_{2}-V_{1})}((1-n)log\tau_{2}-0.577-
(1-n)logt)
\end{equation}
where:
\begin{equation}
\tau_{2}=\tau^{\star}_{\infty}e^{\frac{e_{A}V_{2}}{k_{B}T(1-n)}}
\end{equation}
We have therefore encountered in this limit the ubiquitous 
logarithmic relaxation.
In the literature, the  parameter $S$ is usually defined by:
\begin{equation}
S=\frac{dM(t)/M_{0}}{dlogt}
\end{equation}
which in this case is given by:
\begin{equation}
S=\frac{2k_{B}T(1-n)}{e_{A}(V_{2}-V_{1})}
\end{equation}
Two important comments can now be made. Firstly, the value of $S$ depends on
 $n$ and therefore on the particle interactions and secondly, 
 there are several
possibilities to obtain the logarithmic relaxation: (i) 
from equation (9) and,
assuming that $(1-n)log(t/\tau_{\star})$ is very small, we have\cite{friedrich}:
\begin{equation}
M(t)=M_{0}-2M_{0}e+2(1-n)M_{0}elog(t/\tau_{\star})
\end{equation}
and (ii) assuming a distribution of volume sizes.

\subsection{Distribution of switching fields.}
Following Charap\cite{charap} we consider an ensemble of interacting magnetic
particles with uniaxial anisotropy, thus with a switching barrier
given by equation 7 and a switching field distribution $p(H_{K})$.
We can transform the $p(H_{K})$ in a density of energy barriers
$G(E_{A}^{\star})$ through:
\begin{equation}
G(E_{A}^{\star})=(\frac{\partial E_{A}^{\star}}{\partial
H_{K}})^{-1} p(H_{K})
\end{equation}
In the calculation of $G(E_{A}^{\star})$ we use the Charap method.
For a given $H_{K}$, only for fields $H\approx H_{K}$ we encounter
appreciable switching. In this range of fields we can write
approximately:
\begin{equation}
E_{A}^{\star}\approx\frac{V}{2}M_{s}\frac{(H_{K}-H)^{2}}{H(1-n)}
\end{equation}
which simplifies the evaluation of equation 21.
To take a specific case, we assume a switching field distribution of
Lorentzian form:
\begin{equation}
p(H_{K})=\frac{1}{\pi\Delta(1+(\frac{H_{K}-H_{c}}{\Delta})^{2})}
\end{equation}
where $\Delta$ measures the half-width of the distribution at the
half-maximum.Then $G(E_{A}^{\star})$ can be written as:
\begin{equation}
G(E_{A}^{\star})=\frac{1/\pi\Delta}{ V
M_{s}\sqrt{\frac{2E_{A}^{\star}}{VM_{s}H(1-n)}}(1+(\frac{H-H_{c}+
H(1-n)\sqrt{\frac{2E_{A}^{\star}}{VM_{s}H(1-n)}}}{\Delta})^{2})}
\end{equation}
The magnetization is now described by:
\begin{equation}
\frac{M(t)}{M_{0}}=1-2\int_{0}^{\infty}G(E_{A}^{\star})
e^{-(\frac{t}{\tau_{\star}})^{1-n}}dE_{A}^{\star}
\end{equation}
Defining:
\begin{equation}
\chi=\frac{2E_{A}^{\star}}{VM_{s}H(1-n)}
\end{equation}
\begin{equation}
\beta=\frac{VM_{s}H}{k_{B}T}
\end{equation}
we have:
\begin{equation}
\frac{M(t)}{M_{0}}=1-\frac{H(1-n)}{\pi\Delta}\int_{0}^{\infty}
d\chi\frac{exp(-(\frac{te^{-\beta\chi(1-n)}}{\tau_{\infty}^{\star}})^{1-n}}{\sqrt{\chi}(1+
(\frac{H(1+(1-n)\sqrt{\chi})-H_{c}}{\Delta})^{2})}
\end{equation}
and:
\begin{equation}
S(t)=\frac{H(1-n)^{2}}{\pi\Delta}
(\frac{t}{\tau_{\infty}^{\star}})^{(1-n)}
\int_{0}^{\infty}d\chi
\frac{exp(-\beta\chi(1-n)^{2}-(\frac{t}{\tau_{\infty}^{\star}})^{1-n}.
e^{-\beta\chi(1-n)^{2}})}
{\sqrt{\chi}(1+
(\frac{H(1+(1-n)\sqrt{\chi})-H_{c}}{\Delta})^{2}}
\end{equation}
We have calculated numerically the dependence of $S(t)$ for $t=100s$ with the
applied field at different temperatures for several values of $n$.
A typical example, for $n=0.5$, is shown in figure 3. The choice of
parameters is the same as that of Charap: $\tau_{0}=10^{-9}s$,
$V=10^{-15}cm^{3}$, $H_{c}=200 Oe$, $\Delta=10 Oe$ and 
$t_{c}=10^{-9}s$. Several comments can now be
made. $S_{max}$, the maximum value of $S$, decreases with $n$ and the curve is
broader that the corresponding curve for $n=0$. We have fitted 
the temperature dependence of $S_{max}$
 to a $T^{\alpha}$ law (figure 4) and  have found $\alpha=0.443$. This
does not deviate significantly from  the $n=0$ estimate
$\alpha=0.478$. From a similar calculation given by Charap, we
expect in the first approximation, a dependence of $S_{max}$ with T
of $T^{\frac{1}{2}}$. Finally, we mention that these results are in
qualitative accord with the experimental results of Oseroff el
al~\cite{oseroff}.

\begin{figure}
\vspace{6in}
\caption{Expected dependence of $S$ with $H$ at different 
temperatures according to equation 29}
\end{figure}

\begin{figure}
\vspace{6in}
\caption{$S_{max}$ as function of temperature for n=0.5 and 
fitted to a $T^{\alpha}$ law.}
\end{figure}

\section{Conclusions}
After showing that the Ngai model of relaxation can account for the main
experimental results of the magnetic relaxation in an interacting fine particle
assembly, we have extended the model to the case of a distribution of particles sizes
and switching fields. We mention finally that we look forward to  
this model being used for an analysis of the magnetic 
relaxation in an ensemble of small magnetic particles.

\section{Acknowledgements}
The author is indebted to Dr. D. Fiorani for making available to me  
his experimental results and for many valuable discussions.
Critical reading of the manuscript by Prof. J. Rivas and 
Dr. H. J. Blythe and K. L. Ngai is also
acknowledged.

\end{document}